\documentclass[prd,aps,nofootinbib,onecolumn]{revtex4}
\usepackage{epsfig,graphicx}
\usepackage{amscd}
\usepackage{amsmath}
\usepackage{graphicx}
\usepackage{amssymb}
\usepackage{float}
\usepackage{cancel}
\usepackage{color}

\newcommand{\be}{\begin{eqnarray}}
\newcommand{\ed}{\end{eqnarray}}

\begin{document}
\title{Perturbative QCD study of $B_s$ decays to a pseudoscalar meson and a tensor meson}
\author{Qin Qin$^{1,2}$}
\author{Zhi-Tian Zou$^{3}$}
\author{Xin Yu$^1$}
\author{Hsiang-nan Li$^{4,5,6}$}
\author{Cai-Dian L\"u$^1$}

\affiliation{$^{1}$Institute of High Energy Physics and Theoretical
Physics Center for Science  Facilities, Chinese Academy of Sciences,
Beijing 100049, People's Republic of China,}

\affiliation{$^2$PRISMA Cluster of Excellence $\&$ Mainz Institut
for Theoretical Physics, Johannes Gutenberg University,
Staudingerweg 7, D-55099 Mainz, Germany,}

\affiliation{$^3$Department of Physics, Yantai University, Yantai,
Shandong 264005, People's Republic of China,}

\affiliation{$^4$Institute of Physics, Academia Sinica, Taipei,
Taiwan 115, Republic of China,}

\affiliation{$^5$Department of Physics, National Tsing-Hua
University, Hsinchu, Taiwan 300, Republic of China,}

\affiliation{$^6$Department of Physics, National Cheng-Kung
University, Tainan, Taiwan 701, Republic of China}

\date{\today}

\begin{abstract}
  We study two-body hadronic $B_s\to PT$ decays, with $P (T)$ being a light
  pseudoscalar (tensor) meson, in the perturbative QCD approach.
  The CP-averaged branching ratios and the direct CP asymmetries of the
  $\Delta S=0$ modes are predicted, where $\Delta S$ is the difference between
  the strange numbers of final and initial states. We also define and calculate
  experimental observables for the $\Delta S=1$ modes under the $B_s^0-\bar{B}_s^0$
  mixing, including CP averaged branching ratios, time-integrated CP
  asymmetries, and the CP observables $C_{f}$, $D_{f}$
  and $S_{f}$. Results are compared to the $B_s\to PV$ ones in the literature, and
  to the $B\to PT$ ones, which indicate considerable U-spin symmetry breaking.
  Our work provides theoretical predictions for the $B_s\to PT$
  decays for the first time, some of which will be potentially
  measurable at future experiments.
\end{abstract}


\maketitle

Two-body hadronic $B$ meson decays have attracted a lot of attentions, because of their
importance for studies of CP violation, CKM angle determination, and both weak and
strong dynamics. The two $B$ factories have measured hadronic $B$ decays into
light tensor ($T$) mesons recently \cite{Garmash:2005rv,Aubert:2006fj,Aubert:2008bc},
which were also intensively investigated in several theoretical methods, such as the
naive factorization hypothesis \cite{Katoch:1994zk,Kim:2001sha,Sharma:2010us},
the perturbative QCD (PQCD) approach \cite{Zou:2012td}, and the QCD factorization
approach \cite{Cheng:2010yd}. With much higher production efficiency of $B_s$ mesons
at the LHCb than at the $B$ factories, many data for two-body hadronic $B_s$
decays have been published \cite{Hofer:2010ee,LHCb:2011aa}, but no
decays into tensor mesons were observed so far.

The $B_s$ decays into tensor mesons have not been analyzed theoretically
either to our knowledge. The naive factorization hypothesis does not apply to
modes involving only the annihilation amplitudes, and only
the amplitudes with tensor mesons being emitted from the weak vertex. Besides,
branching ratios for color-suppressed decays estimated in the naive factorization
are usually too small. As for the QCD factorization\cite{Beneke:1999br}, owning to lack of data for
$B_s\to PT$ branching ratios, $P$ being a light pseudoscalar meson, the
penguin-annihilation parameters cannot be determined through global fits. If the
parameters associated with the $B_s\to PT$ modes were approximated by the
$B_s\to PV$ ones \cite{Cheng:2010yd}, large theoretical
uncertainties would be introduced. Both the annihilation amplitudes and
the nonfactorizable tensor-emission amplitudes are calculable in the
PQCD approach without inputs of free parameters. Encouraged by successful applications
of the PQCD approach to many two-body hadronic $B$ meson decays
\cite{KLS,Ali:2007ff,Zou:2012td,Lu:pqcd}, we will make predictions for the $B_s\to PT$
branching ratios and CP-violation observables in this letter, which
can provide useful hints to relevant experiments.

The effective electroweak Hamiltonian relevant to the $B_s\to PT$ decays is written as
\begin{equation}\label{hamiltonian}
\mathcal{H}_{eff}={G_F\over\sqrt{2}}\left[\sum_{i=1}^2V^*_{ub}V_{uD}C_i(\mu)
O_i^u(\mu)-\sum_{j=3}^{10}V^*_{tb}V_{tD}C_j(\mu)O_j^u(\mu)\right],
\end{equation}
where $V$'s are the CKM matrix elements with $D$ denoting a down-type quark $d$ or $s$,
$O_{i,j}(\mu)$ are the tree and penguin four-quark operators \cite{Buras:ewet}, and
$C_{i,j}(\mu)$ are the corresponding
Wilson coefficients, which evolve from the $W$ boson mass down to the renormalization
scale $\mu$. In the PQCD approach a hadronic transition matrix element
of a four-quark operator is further factorized into two pieces \cite{Chang:1996dw}:
the kernel with hard gluon exchanges characterized by the $b$ quark mass,
and the nonperturbative hadron wave functions characterized by the QCD scale
$\Lambda_{\rm QCD}$.

\begin{figure}[!ht]
  \begin{center}
  \includegraphics[scale=0.5]{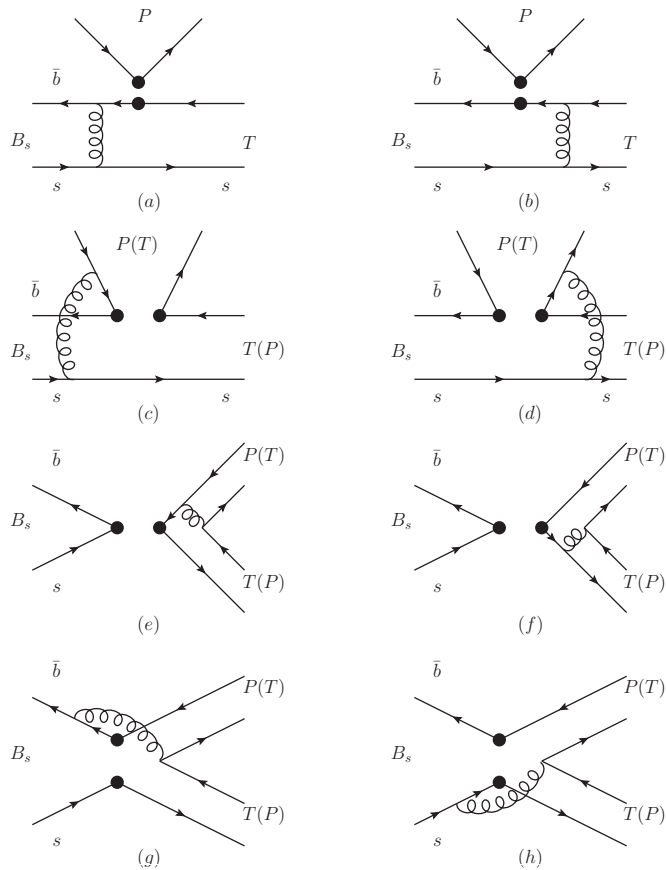}
  \vspace{-1cm}
  \caption{Leading-order diagrams for $B_s\to PT$ decays.
}\label{diagrams}
  \end{center}
\end{figure}

The leading-order diagrams contributing to the $B_s\to PT$ decays are displayed in
Fig.~\ref{diagrams}, where (a) and (b) are factorizable emission-type diagrams,
(c) and (d) are nonfactorizable emission-type diagrams, (e) and (f) are
factorizable annihilation-type diagrams, and (g) and (h) are nonfactorizable
annihilation-type diagrams. As indicated in Fig.~\ref{diagrams}, the factorizable
tensor-emission amplitudes do not exist, since a tensor meson cannot be produced via
a $V$ or $A$ current. The PQCD results for the $B\to PT$ (without $B_s$) decays
\cite{Zou:2012td} are basically in agreement with the experimental data
\cite{Beringer:1900zz,Amhis:2012bh} and those from the QCD factorization
\cite{Cheng:2010yd}. The extension of the PQCD formalism to the $B_s\to PT$
decays is straightforward because of the similarity between $B$ and $B_s$
decays in SU(3) symmetry: the factorization formula for every diagram can be
obtained by substituting the quantities in the $B_s\to PT$ decays for the
corresponding ones in the $B\to PT$ decays \cite{Zou:2012td}. The confrontation
of the $B\to PT$ calculations to the data has restricted the parameters involved
in the $P$ and $T$ meson wave functions to some extent. In this work we will adopt
the $B_s$ meson wave function in \cite{Ali:2007ff}, and the $P$ and $T$
meson wave functions in \cite{Zou:2012td}.

A neutral meson and its charge conjugate partner, including the $K^0-\bar{K}^0$,
$D^0-\bar{D}^0$, $B^0-\bar{B}^0$, and $B^0_s-\bar{B}^0_s$ systems, mix through
the weak interaction. The $B^0_s-\bar{B}^0_s$ mixing is the strongest,
since the mass difference $\Delta M$ between the mass eigenstates is much larger than
the decay width $\Gamma$ of the $B_s$ meson. The frequent oscillation between
the $B_s^0$ and $\bar{B_s^0}$ mesons due to the strong mixing has rendered difficult
measurements of $B_s$ decay observables at the $B$ factories, such as measurements
of time-dependent CP-violation parameters. However, these measurements become
feasible in LHCb experiments, because of the time dilation caused by energetic
$B_s$ mesons. The mass eigenstates of the $B_s$ mesons are superpositions of the
flavor eigenstates,
\begin{equation}\label{superposition}
|B_{sL,H}\rangle=p|B_s^0\rangle\pm q|\bar{B}_s^0\rangle,
\end{equation}
where $p$ and $q$ are complex coefficients. We neglect the difference between the mass
eigenstates and the CP eigenstates, and assume that $B_{sL(H)}$ is CP even (odd)
as suggested in \cite{Lenz:2006hd}. The time-dependent $B_s\to PT$ differential
branching ratios are then expressed as \cite{neutralCPA}
\begin{equation}\begin{split}\label{diff-br}
\frac{d}{dt}Br(B_s^0(t)\to f)&=\Phi(B_s\to f)e^{-{\Gamma}t}|A_f|^2{1+|\lambda_f|^2\over2}\\
&\times\left[\cosh({\Delta\Gamma\over2}t)+\cos({\Delta M}t)C_f
-\sin({\Delta M}t)S_f-\sinh({\Delta\Gamma\over2}t)D_f\right],\\
\frac{d}{dt}Br(\bar{B}_s^0(t)\to f)&=\Phi(B_s\to f)
e^{-{\Gamma}t}|{p\over q}|^2|A_f|^2{1+|\lambda_f|^2\over2}\\
&\times\left[\cosh({\Delta\Gamma\over2}t)-\cos({\Delta M}t)C_f
+\sin({\Delta M}t)S_f-\sinh({\Delta\Gamma\over2}t)D_f\right],\\
\frac{d}{dt}Br(\bar{B}_s^0(t)\to \bar{f})&=\Phi(B_s\to f)
e^{-{\Gamma}t}|\overline{A}_{\overline{f}}|^2{1+|\overline{\lambda}_{\overline{f}}|^2\over2}\\
&\times\left[\cosh({\Delta\Gamma\over2}t)+\cos({\Delta M}t)C_{\overline{f}}
-\sin({\Delta M}t)S_{\overline{f}}-\sinh({\Delta\Gamma\over2}t)D_{\overline{f}}\right],\\
\frac{d}{dt}Br(B_s^0(t)\to \bar{f})&=\Phi(B_s\to f)e^{-{\Gamma}t}|
{q\over p}|^2|\overline{A}_{\overline{f}}|^2{1+|\overline{\lambda}_{\overline{f}}|^2\over2}\\
&\times\left[\cosh({\Delta\Gamma\over2}t)-\cos({\Delta M}t)C_{\overline{f}}
+\sin({\Delta M}t)S_{\overline{f}}-\sinh({\Delta\Gamma\over2}t)D_{\overline{f}}\right],
\end{split}\end{equation}
with the mass difference $\Delta M=(116.4\pm0.5)\times10^{-10}$ MeV,
the decay width difference $\Delta\Gamma=(0.100\pm0.013)\times10^{12}$ $s^{-1}$
\cite{Beringer:1900zz}, $\Phi(B_s\to f)$ being the phase space of the corresponding
mode, and $A_f$ ($\bar{A}_{\bar{f}}$) being the $B_s^0\to f$ ($\bar{B}_s^0\to\bar{f}$)
decay amplitude. We have employed the definitions of the amplitude ratios $\lambda_f$
and $\bar{\lambda}_{\bar{f}}$, and the CP asymmetry observables $C_{f,\bar{f}}$,
$D_{f,\bar{f}}$ and $S_{f,\bar{f}}$ used in \cite{neutralCPA}.

Since the oscillation
period is much shorter than the lifetime of the $B_s$ meson, Eq.~(\ref{diff-br}) can
be integrated over $t$, and lead to the time-integrated branching ratios
\begin{equation}\begin{split}\label{inte-br}
Br(B_s^0(\infty)\to f)&=\Phi(B_s\to f)|A_f|^2{1+|\lambda_f|^2\over2}
\left[{\Gamma-D_f{\Delta\Gamma\over2}\over \Gamma^2}
+{C_f\Gamma+S_f\Delta M\over \Gamma^2+\Delta M^2}\right],\\
Br(\bar{B}_s^0(\infty)\to f)&=\Phi(B_s\to f)|A_f|^2{1+|\lambda_f|^2\over2}
\left[{\Gamma-D_f{\Delta\Gamma\over2}\over \Gamma^2}-{C_f\Gamma
+S_f\Delta M\over \Gamma^2+\Delta M^2}\right],\\
Br(\bar{B}_s^0(\infty)\to \bar{f})&=\Phi(B_s\to f)|\overline{A}_{\overline{f}}|^2
{1+|\overline{\lambda}_{\overline{f}}|^2\over2}
\left[{\Gamma-D_{\bar{f}}{\Delta\Gamma\over2}\over \Gamma^2}
+{C_{\bar{f}}\Gamma+S_{\bar{f}}\Delta M\over\Gamma^2+\Delta M^2}\right],\\
Br(B_s^0(\infty)\to \bar{f})&=\Phi(B_s\to f)|\overline{A}_{\overline{f}}|^2
{1+|\overline{\lambda}_{\overline{f}}|^2\over2}
\left[{\Gamma-D_{\bar{f}}{\Delta\Gamma\over2}\over \Gamma^2}
-{C_{\bar{f}}\Gamma+S_{\bar{f}}\Delta M\over\Gamma^2+\Delta M^2}\right].
\end{split}\end{equation}
The terms proportional to $(\Delta\Gamma/\Gamma)^2 \approx 0.006$ have been
dropped, and the approximation $|p/ q|^2=1$ has been made in the above
expressions. If it happens that the $B_{sL}$ state is CP odd while $B_{sH}$
is CP even, the substitutions $\Delta M\to-\Delta M$ and
$\Delta\Gamma\to-\Delta\Gamma$, or
equivalently, $D_{f,\bar{f}}\to-D_{f,\bar{f}}$ and
$S_{f,\bar{f}}\to-S_{f,\bar{f}}$ need to be done.

For $\Delta S=0$ modes, a $B_s^0$ ($\bar{B}_s^0$) meson decays to the final
state $f$ ($\bar{f}$), but not to $\bar{f}$ ($f$) with $f\neq\bar{f}$. In
this case one can determine the initial $B_s^0$ or $\bar{B}_s^0$
meson through the final state even under the frequent $B_s^0-\bar{B}_s^0$
oscillation. The ordinary definitions of CP-averaged branching ratios and
direct CP asymmetries then apply directly. The predictions for the
CP-averaged branching ratios and the direct CP asymmetries of these
$\Delta S=0$ modes are listed in Table~\ref{br1}. The dominant topological
amplitudes for each decay channel are also listed, including the color-favored
($T$), color-suppressed ($C$), and annihilation-type ($A$) tree amplitudes, and
the corresponding penguin amplitudes $PT$, $PC$, and $PA$. Two types of
theoretical uncertainties are estimated here: the first type comes from the
variation of the nonperturbative parameters in the meson wave functions
(see \cite{Ali:2007ff,Zou:2012td}, except that we have adopted the recent
lattice QCD result for the $B_s$ meson decay constant, 0.228(10) GeV \cite{Na:2012kp});
the second type reflects the unknown next-to-leading-order QCD corrections
characterized by the variations of the QCD scale $\Lambda_{\rm QCD}=(0.25\pm0.05)$
GeV and of the hard scales. It is observed that both types of uncertainties are
roughly of the same order for most channels.

As shown in Table~\ref{br1}, only the $B_s^0\to\pi^+K_2^{*-}$ decay has
a sizable branching ratio arising from the dominant amplitude $T$, and the
branching ratios of the other modes are of order $10^{-7}$. For
color-suppressed modes such as $B_s^0\to\bar{K}^0a_2^0$, $\bar{K}^0f_2$ and
$\bar{K}^0f_2'$, there is no significance difference between their branching
ratios and those of their $PV$ partners \cite{Ali:2007ff}, because the
factorizable emission contributions are less important. For the color-favored
$B_s^0\to K^-a_2^+$ decay, whose factorizable tensor-emission amplitude is
forbidden, its branching ratio $1.50\times10^{-7}$ is much smaller than the
$B_s^0\to K^-\rho^+$ one, $1.78\times10^{-5}$. Most modes in Table~\ref{br1}
exhibit large direct CP asymmetries caused by the interference between the
tree and penguin amplitudes. The direct CP asymmetry in the
$B_s^0\to \bar{K}^0f_2'$ decay would vanish, if $f_2'$ was a pure $\bar{s}s$
state. After receiving a tree contribution from the mixing of the isospin-1
states, this mode gets a small CP asymmetry.

To examine whether the U-spin symmetry holds in the $B_{(s)}\to PT$ decays, we
define the following ratios
\begin{equation}\begin{split}
&R_{CP}(B_s^0\to f)\equiv-{A_{CP}(B_s^0\to f)\over A_{CP}(B^0\to Uf)},\\
&R_{\Gamma}(B_s^0\to f)\equiv{\tau(B_s^0)\over\tau(B^0)}{Br(B^0\to Uf)\over Br(B_s^0\to f)},
\end{split}\end{equation}
where $U$ stands for the U-spin transformation, $d\leftrightarrow s$.
The relation between two decay modes in a U-spin pair implies that the above
ratios are equal to each other in the U-spin symmetry limit \cite{Gronau:2013mda}.
Combing our predictions with the $B\to PT$ ones \cite{Zou:2012td}, we obtain
$R_{CP}(B_s^0\to\pi^+K_2^{*-})=0.29^{+0.10}_{-0.08}$ and
$R_{\Gamma}(B_s^0\to\pi^+K_2^{*-})=0.74^{+0.24}_{-0.19}$;
$R_{CP}(B_s^0\to K^-a_2^+)=1.9^{+0.5}_{-0.5}$ and
$R_{\Gamma}(B_s^0\to K^-a_2^+)=5.2^{+0.9}_{-0.6}$.
The central values indicate that the U-spin symmetry is considerably broken in
the $B_{(s)}\to PT$ decays by hadronic effects at order
$(m_s-m_d)/\Lambda_{\rm QCD}$ \cite{Gronau:2013mda}, $m_s$ ($m_d$) being the
strange (down) quark mass. The physical U-spin conjugate
processes of the other modes do not exist due to the superposition of the
flavor states $\bar{q}q$ in final-state mesons.

\begin{table}[!htbh]
  \centering
  \caption{Branching ratios (in units of $10^{-7}$) and direct CP asymmetries of
  the $\Delta$S=0 $B_s^0\to PT$ decays.}\label{br1}
  \begin{tabular}[t]{cccc}\hline\hline
  Modes & Amplitudes & $Br$ & Direct $A_{CP}$ ($\%$)\\\hline
  $B_s^0\to \pi^+K_2^{*-}$ & $T$ & $90^{+40+4}_{-32-6}$ & $13^{+2+2}_{-2-2}$\\
  $B_s^0\to \pi^0\bar{K}_2^{*0}$ & $C$,$PA$ & $1.3^{+0.6+0.6}_{-0.5-0.5}$ & $47^{+8+9}_{-6-6}$ \\
  $B_s^0\to \bar{K}^0a_2^0$ & $C$,$PA$ & $2.0^{+0.4+0.2}_{-0.3-0.3}$ & $38^{+7+6}_{-10-7}$ \\
  $B_s^0\to \bar{K}^0f_2$ & $C$,$PA$ & $3.4^{+0.7+0.7}_{-0.6-0.7}$ & $-24^{+5+3}_{-6-5}$ \\
  $B_s^0\to \bar{K}^0f_2'$ & $PA$ & $2.0^{+0.5+0.8}_{-0.4-0.6}$ & $4.8^{+2.8+1.9}_{-1.7-1.4}$ \\
  $B_s^0\to K^-a_2^+$ & $T$,$PA$ & $1.5^{+0.3+0.4}_{-0.2-0.3}$ & $39^{+8+1}_{-1-4}$ \\
  $B_s^0\to \eta\bar{K}_2^{*0}$ & $C$,$PA$ & $0.55^{+0.29+0.35}_{-0.19-0.27}$ & $77^{+13+5}_{-12-2}$ \\
  $B_s^0\to \eta'\bar{K}_2^{*0}$ & $C$,$PT$ & $3.5^{+1.2+1.4}_{-1.0-1.2}$ & $-30^{+2+7}_{-1-6}$  \\
  \hline\hline
  \end{tabular}
\end{table}

For $\Delta S=1$ $B_s^0$ ($\bar{B}_s^0$) meson decays, we first consider those
modes, whose final states are CP eigenstates, i.e. $f=\bar{f}$.
In this case the four equations in Eq.~(\ref{diff-br}) reduce to two, and
one has to measure the CP observables $C_f$, $D_f$ and $S_f$ through
time-dependent branching ratios, which require a lot of data accumulation.
Alternatively, we define the time-integrated CP asymmetries for these decays
\begin{equation}\begin{split}\label{ti_in_CPA}
A_{CP}(B_s(\infty)\to f)&\equiv{Br(\bar{B}_s^0(\infty)\to f)-Br(B_s^0(\infty)\to f)
\over Br(\bar{B}_s^0(\infty)\to f)+Br(B_s^0(\infty)\to f)}\\
&=-{C_f\Gamma+S_f\Delta M\over \Gamma^2+\Delta M^2}
{\Gamma^2\over\Gamma-D_f{\Delta\Gamma\over2}},
\end{split}\end{equation}
and assess if there is a chance to measure it at the early stage of
data accumulation.

The PQCD predictions for all the experimental observables,
together with the dominant topological amplitudes and uncertainties,
are shown in Table~\ref{br2}. It is observed that the $\eta'$-involved modes
$B_s^0\to\eta' a_2^0(f_2,f_2')$ have branching ratio larger than those of the
corresponding $\eta$-involved modes $B_s^0\to\eta a_2^0(f_2,f_2')$. This
pattern is understood, since the dominant amplitudes require the
$\bar{s}s$ constituent, which is more in $\eta'$ than in $\eta$.
The branching ratios of the $\Delta I=1$ modes, like
$B_s^0\to\eta a_2^0$ and $\eta' a_2^0$, are highly suppressed, compared
to those of the corresponding $\Delta I=0$ modes, $B_s^0\to\eta f_2$ and
$\eta' f_2$. This suppression can be explained as follows. Neglecting the
$f_2-f_2'$ mixing effect, both $B_s^0\to\eta'a_2^0$ and $\eta'f_2$
are dominated by the amplitudes $PC$ naively. However, the minus sign
in the flavor constituent $(\bar{u}u-\bar{d}d)/\sqrt{2}$ renders
$PC(u)$ and $PC(d)$ cancel in the former mode, while they become constructive
in the latter. The source of the discrepancy between
the $B_s^0\to\eta a_2^0$ and $\eta f_2$ branching ratios is the same.

Contrary to the $\Delta S=0$ decays, the tree and penguin contributions
are never simultaneously sizable to form significant interferences in the
$\Delta S=1$ decays listed in Table~\ref{br2}, so the direct CP violation
$C_f$'s are tiny. One seemingly exceptional mode is $B_s^0\to\pi^0f_2$,
which has the tree and penguin contributions of the same order, but still a
small direct CP asymmetry. A careful investigation reveals that the strong
phases of the tree and penguin amplitudes are almost equal,
$\phi^s_T\approx\phi^s_P$, and the direct CP asymmetry is proportional to
$\sin(\phi^s_T-\phi^s_P)$ \cite{Isidori:2011qw}. Besides, the
time-integrated CP asymmetries in Table~\ref{br2} differ dramatically
from the corresponding direct CP asymmetries $-C_f$'s. According to
Eq.~(\ref{ti_in_CPA}), the differences mainly come from the large mixing
parameter $\Delta M$.

\begin{table}[!htbh]
  \centering
  \caption{Branching ratios (in units of $10^{-7}$) and CP observables for
  the $\Delta S=1$ $B_s^0\to PT$ decays, whose final states are CP eigenstates.}\label{br2}
  \begin{tabular}[t]{ccccccc}\hline\hline
  Modes & Amplitudes & $Br$ & $C_f$ & $D_f$ & $S_f$ & time-inte $A_{CP} ($\%$)$\\\hline
  $\pi^0a_2^0$ & $PA$ & $0.90^{+0.19+0.31}_{-0.14-0.31}$ & $-0.082^{+0.072+0.055}_{-0.001-0.015}$ & $-0.988^{+0.003+0.001}_{-0.003-0.003}$ & $-0.133^{+0.021+0.008}_{-0.031-0.011}$ & $0.50^{+0.10+0.03}_{-0.10-0.03}$\\
  $\pi^0f_2$ & $A$,$PC$ & $0.048^{+0.012+0.002}_{-0.016-0.012}$ & $-0.04^{+0.06+0.02}_{-0.12-0.06}$ & $-0.66^{+0.08+0.08}_{-0.02-0.04}$ & $0.75^{+0.06+0.06}_{-0.01-0.04}$ & $-2.7^{+0.1+0.2}_{-0.2-0.2}$\\
  $\pi^0f_2'$ & $PC$ & $1.2^{+0.6+0.1}_{-0.5-0.1}$ & $-0.05^{+0.01+0.01}_{-0.02-0.02}$ & $-0.95^{+0.01+0.03}_{-0.01-0.02}$ & $0.30^{+0.03+0.07}_{-0.02-0.07}$ & $-1.0^{+0.1+0.3}_{-0.1-0.3}$\\
  $\eta a_2^0$ & $C$,$A$ & $0.047^{+0.013+0.010}_{-0.010-0.012}$ & $0.02^{+0.01+0.01}_{-0.02-0.06}$ & $0.40^{+0.01+0.06}_{-0.01-0.04}$ & $0.92^{+0.01+0.02}_{-0.01-0.03}$ & $-3.6^{+0+0.1}_{-0-0.1}$\\
  $\eta f_2$ & $PC$ & $9.8^{+2.7+3.2}_{-2.2-2.6}$ & $-0.014^{+0.003+0.008}_{-0.008-0.010}$ & $-0.995^{+0.001+0.002}_{-0.001-0}$ & $-0.098^{+0.007+0.004}_{-0.007-0.020}$ & $0.30^{+0.02+0.07}_{-0.02-0.01}$\\
  $\eta f_2'$ & $PA$ & $96^{+20+36}_{-19-30}$ & $0.022^{+0.004+0.003}_{-0.004-0.003}$ & $-1.000^{+0+0}_{-0-0}$ & $0.024^{+0.004+0.003}_{-0.004-0.005}$ & $-0.10^{+0.02+0.02}_{-0.01-0.01}$\\
  $\eta' a_2^0$ & $C$,$A$ & $0.13^{+0.03+0.03}_{-0.03-0.03}$ & $0.03^{+0.01+0.02}_{-0.01-0.01}$ & $0.28^{+0.03+0.04}_{-0-0.03}$ & $0.96^{+0+0.01}_{-0.01-0.01}$ & $-3.7^{+0+0.1}_{-0-0}$\\
  $\eta' f_2$ & $PC$ & $30^{+7+11}_{-7-10}$ & $-0.005^{+0+0.002}_{-0.012-0.010}$ & $-0.994^{+0.001+0.001}_{-0.001-0.001}$ & $-0.104^{+0.011+0.006}_{-0.006-0.006}$ & $0.40^{+0.02+0.02}_{-0.04-0.02}$\\
  $\eta' f_2'$ & $PA$,$PT$ & $245^{+69+99}_{-59-84}$ & $-0.007^{+0.004+0}_{-0.003-0.001}$ & $-1.000^{+0+0}_{-0-0}$ & $-0.009^{+0.006+0.004}_{-0.002-0.001}$ & $0.030^{+0.010+0.002}_{-0.020-0.010}$\\
  \hline\hline
  \end{tabular}
\end{table}

There exist more complicated $\Delta S=1$ modes, in which either a $B_s^0$ or
$\bar{B}_s^0$ meson can decay into $f$ and $\bar{f}$ with $f\neq\bar{f}$.
Even though a final state is identified in this case, there is no way to
determine whether the initial state is
a $B_s^0$ or $\bar{B}_s^0$ meson directly. It is then difficult to
distinguish the four channels in Eq.~(\ref{diff-br}), and time-dependent
measurements are also required. For experimental access,
we define the CP asymmetry parameter only by charge-tag of final states
\begin{equation}\begin{split}\label{ti_fla}
A_{CP}&\equiv{Br(B_s^0/\bar{B}^0_s(\infty)\to \bar{f})-Br(B_s^0/\bar{B}^0_s(\infty)\to f)\over
Br(B_s^0/\bar{B}^0_s(\infty)\to \bar{f})+Br(B_s^0/\bar{B}^0_s(\infty)\to f)}.
\end{split}\end{equation}
All the CP observables, and the sum of the branching ratios of a pair of
channels defined by
\begin{equation}
Br\equiv\frac{1}{2}\left[Br(B_s^0(\infty)\to f)+Br(\bar{B}_s^0(\infty)\to\bar{f})
+Br(B_s^0(\infty)\to\bar{f})+Br(\bar{B}_s^0(\infty)\to f)\right],
\end{equation}
are presented in Table~\ref{br4}. For the
$B_s^0\to\bar{K}^0K_2^{*0}$ set, all the $f$-related CP observables are
equal to the $\bar{f}$-related ones, and the CP asymmetry parameter
$A_{CP}$ is exactly zero. There are no tree
contributions, and the penguin amplitudes share one common weak phase in
these decays. It is then straightforward to arrive at
$\lambda_f=\bar{\lambda}_{\bar{f}}$, and thus
$C(D,S)_f=C(D,S)_{\bar{f}}$ and $A_{CP}=0$.

\begin{table}[!htbh]
  \centering
  \caption{Branching ratios (in units of $10^{-7}$) and CP observables for the rest $\Delta S=1$ decays.}\label{br4}
  \begin{tabular}[t]{ccccccccc}\hline\hline
  Modes & $C_f$ & $D_f$ & $S_f$ & $C_{\bar{f}}$ & $D_{\bar{f}}$ & $S_{\bar{f}}$ & $Br$ & $A_{CP} ($\%$)$ \\\hline
  $\pi^+a_2^-$  & $-0.15^{+0.01+0.02}_{-0.04-0.05}$ & $-0.98^{+0+0.01}_{-0-0.01}$ & $-0.10^{+0.07+0.05}_{-0.01-0.01}$  & $-0.05^{+0.07+0.07}_{-0.02-0.01}$ & $-0.98^{+0.01+0.01}_{-0.01-0.01}$ & $0.18^{+0.04+0.04}_{-0.02-0.03}$ & $1.8^{+0.4+0.6}_{-0.2-0.8}$ & $13^{+3+5}_{-5-5}$\\
  $K^+K_2^{*-}$  & $0.49^{+0.07+0.02}_{-0.06-0.01}$ & $-0.85^{+0.04+0}_{-0.03-0}$ & $-0.18^{+0.02+0.03}_{-0.04-0.05}$  & $0.03^{+0.11+0.09}_{-0.08-0.13}$ & $-0.71^{+0.09+0.03}_{-0.06-0.02}$ & $-0.70^{+0.07+0.03}_{-0.07-0.03}$ & $86^{+20+28}_{-16-24}$ & $-28^{+2+5}_{-3-6}$\\
  $K^0\bar{K}_2^{*0}$  & $0.24^{+0.08+0.03}_{-0.06-0.05}$ & $-0.91^{+0.03+0.02}_{-0.02-0.02}$ & $-0.34^{+0.03+0.04}_{-0.03-0.03}$ & $0.24^{+0.08+0.03}_{-0.06-0.05}$ & $-0.91^{+0.03+0.02}_{-0.02-0.02}$ & $-0.34^{+0.03+0.04}_{-0.03-0.03}$ & $70^{+14+24}_{-12-20}$ & 0\\
  \hline\hline
  \end{tabular}
\end{table}

In this letter we have investigated the $B_s\to PT$ decays in the PQCD approach,
whose branching ratios and CP asymmetry parameters were predicted. It was
noticed that the absence of the factorizable tensor-emission amplitudes in these
decays leads to differences from the $B_s\to PV$ ones.
Owing to the significant $B_s^0-\bar{B}_s^0$ mixing effect, the time-integrated
CP asymmetries have been redefined and calculated for the $\Delta S=1$
modes. The U-spin symmetry was found to be considerably broken, when
the $B_s^0\to\pi^+ K_2^{*-}$ and $K^-a_2^+$ branching ratios are compared
to the corresponding $B^0\to K^+a_2^-$ and $\pi^-K_2^{*+}$ ones.
The branching ratios of some modes reach $\mathcal{O}(10^{-6})$
or even $\mathcal{O}(10^{-5})$, including $B_s^0\to\eta f_2'$, $\eta'f_2$,
$\eta'f_2'$, $K^+K_2^{*-}$, $K^0\bar{K}_2^{*0}$, and $\pi^+K_2^{*-}$, which
are expected to be measured at LHCb experiments. There is
also potential to observe CP violation effects in the $B_s^0\to\pi^+K_2^{*-}$,
$K^+K_2^{*-}$ and $K^0\bar{K}_2^{*0}$ decays in the near future.

We are grateful to Prof. Yuan-Ning Gao and Wen-Fei Wang for helpful discussions,
and to Prof. Matthias Neubert for useful comments and suggestions.
The research of Q.~Qin was also supported in part by the CAS-DAAD Joint
Fellowship Programme under PKZ A1394070 and the Cluster of Excellence {\em
Precision Physics, Fundamental Interactions and Structure of Matter\/}
(PRISMA -- EXC 1098). The research of Z.-T. Zou was supported in part by
the Foundation of Yantai University under Grant No. WL07052. This work was
also partially supported by the National Science Foundation of China under
Grant Nos. 11375208, 11228512 and 11235005, and by the National Science
Council of R.O.C. under Grant No. NSC-101-2112-M-001-006-MY3.

\end{document}